\documentstyle[12pt,epsf]{article}
 \newcommand{\insertplot}[5]{\begin{figure}
 \hfill\hbox to 0.05in{\vbox to #5in{\vfill
 \inputplot{#1}{#4}{#5}}\hfill}
 \hfill\vspace{-.1in}
 \caption{#2}\label{#3}
 \end{figure}}
 \newcommand{\inputplot}[3]{
 \special{ps: plotfile #1}
}

\begin{document}

\title{SU(3) Einstein-Yang-Mills-Dilaton Sphalerons and Black Holes}
\vspace{1.5truecm}
\author{
{\bf Burkhard Kleihaus$^1$, Jutta Kunz$^{1,2}$ and Abha Sood$^1$}\\
$^1$Fachbereich Physik, Universit\"at Oldenburg, Postfach 2503\\
D-26111 Oldenburg, Germany\\
$^2$
Instituut voor Theoretische Fysica, Rijksuniversiteit te Utrecht\\
NL-3508 TA Utrecht, The Netherlands}


\maketitle
\vspace{1.0truecm}

\begin{abstract}
SU(3) Einstein-Yang-Mills-dilaton theory possesses sequences of
static spherically symmetric sphaleron
and black hole solutions for the SU(2) and the SO(3) embedding.
The solutions depend on the dilaton coupling constant $\gamma$,
approaching the corresponding
Einstein-Yang-Mills solutions for $\gamma \rightarrow 0$,
and Yang-Mills-dilaton solutions in flat space for $\gamma \rightarrow \infty$.
The sequences of solutions tend to Einstein-Maxwell-dilaton
solutions with different magnetic charges.
The solutions satisfy analogous relations
between the dilaton field and the metric for general $\gamma$.
Thermodynamic properties of the black hole solutions are discussed.
\end{abstract}
\vfill
 \noindent {Preprint hep-th/9510069} \hfill\break
\vfill\eject

\section{Introduction}

SU(2) Yang-Mills theory in $3+1$ dimensions
does not possess static, finite energy solutions.
In contrast, SU(2) Einstein-Yang-Mills theory
possesses a sequence of static, spherically symmetric
sphaleron solutions \cite{bm,strau1,volkov1},
and so does SU(2) Yang-Mills-dilaton theory \cite{lav1,bizon2}
as well as SU(2) Yang-Mills-Higgs theory,
where the electroweak sphaleron \cite{dhn,bog,km}
is accompanied by such a sequence
for large Higgs boson masses
\cite{kb1,yaffe}.

Motivated by higher-dimensional unified theories,
such as Kaluza-Klein theory or superstring theory,
where a scalar dilaton field arises naturally,
recently SU(2) Einstein-Yang-Mills-dilaton theory
has been studied and shown to also
possess a sequence of static spherically symmetric
regular particle-like solutions
which are unstable \cite{don,lav2,maeda,bizon3,neill}.

Like SU(2) Einstein-Yang-Mills theory
\cite{volkov,bizon1,kuenzle1},
SU(2) Einstein-Yang-Mills-dilaton theory
additionally possesses a sequence of
static spherically symmetric black hole solutions
with non-trivial matter fields
\cite{don,lav2,maeda,neill}.
The sequences of black hole solutions are unstable, too
\cite{strau2,volkov5,volkov4,lav2}.

Here we consider
SU(3) Einstein-Yang-Mills-dilaton theory.
The equations of motion obtained for
static spherically symmetric ans\"atze
for the metric and for the matter fields
(with magnetic gauge field)
yield relations between the dilaton field
and the metric functions.
We construct sphaleron and black hole solutions,
employing both the
SU(2) and the SO(3) embedding.
The solutions are similar to the corresponding solutions
of SU(3) Einstein-Yang-Mills theory \cite{kuenzle,kks},
and approach these for small dilaton coupling constant $\gamma$.

Labeling the solutions by the number of nodes
of the gauge field functions, $n$,
we construct the sphaleron and black hole solutions
with four or less nodes.
We relate the properties of the regular and black hole
solutions in the limit $n \rightarrow \infty$
to those of black hole solutions of
Einstein-Maxwell-dilaton theory \cite{gib1,str}.

\section{SU(3) Einstein-Yang-Mills-Dilaton Equations of Motion}

We consider the SU(3) Einstein-Yang-Mills-dilaton action
\begin{equation}
S=S_G+S_M=\int L_G \sqrt{-g} d^4x + \int L_M \sqrt{-g} d^4x
\   \end{equation}
with
\begin{equation}
L_G=\frac{1}{16\pi G}R
\ , \end{equation}
and the matter Lagrangian
\begin{equation}
L_M=-\frac{1}{2}\partial_\mu \Phi \partial^\mu \Phi
 -e^{2 \kappa \Phi }\frac{1}{2} {\rm Tr} (F_{\mu\nu} F^{\mu\nu})
\ , \end{equation}
where
\begin{equation}
F_{\mu\nu}= \partial_\mu A_\nu - \partial_\nu A_\mu
            - i g [A_\mu,A_\nu]
\ , \end{equation}
\begin{equation}
A_\mu = \frac{1}{2} \lambda^a A_\mu^a
\ , \end{equation}
and $g$, $\kappa$ are the gauge and dilaton coupling constants,
respectively.
Variation of the action eq.~(1) with respect to the metric
$g_{\mu\nu}$ leads to the Einstein equations,
and variation with respect to the gauge field $A_\mu$
and the dilaton field $\Phi$
to the matter field equations.

To construct static spherically symmetric
regular and black hole solutions
we employ Schwarz\-schild-like coordinates and adopt
the spherically symmetric metric
\begin{equation}
ds^2=g_{\mu\nu}dx^\mu dx^\nu=
  -A^2N dt^2 + N^{-1} dr^2 + r^2 (d\theta^2 + \sin^2\theta d\phi^2)
\ , \label{metric} \end{equation}
with
\begin{equation}
N=1-\frac{2m}{r}
\ . \end{equation}
Generalized spherical symmetry for the gauge field is realized
by embedding the SU(2) or the SO(3) generators in SU(3).
In the SU(2)-embedding
the ansatz for the gauge field
with vanishing time component is \cite{bm},
\begin{eqnarray}
A_0 &=& 0
\ , \nonumber\\
A_i &=& \frac{1-w(r)}{2rg} (\vec e_r \times \vec \tau)_i
\ , \label{su2} \end{eqnarray}
with the SU(2) Pauli matrices
$\vec \tau = (\tau_1,\tau_2,\tau_3)$.
In the SO(3)-embedding
the corresponding ansatz for the gauge field
with vanishing time component is
\begin{eqnarray}
A_0 &=& 0
\ , \nonumber\\
A_i &=& \frac{2-K(r)}{2rg} (\vec e_r \times \vec \Lambda)_i
       +\frac{H(r)}{2rg} \Bigl[ (\vec e_r \times \vec \Lambda)_i,
       \vec e_r \cdot \vec \Lambda) \Bigr]_+
\ , \label{so3} \end{eqnarray}
where $[\ , \ ]_+$ denotes the anticommutator, and
$\vec \Lambda = (\lambda_7,-\lambda_5,\lambda_2)$.
For the dilaton field we take $\Phi = \Phi(r)$.

The SU(2)-embedding, eq.~(\ref{su2}),
leads to the previously studied
SU(2) Einstein-Yang-Mills-dilaton equations
\cite{don,lav2,maeda,bizon3,neill}.
To obtain the SU(3) Einstein-Yang-Mills-dilaton equations
for the SO(3)-embed\-ding, eq.~(\ref{so3}),
we employ the $tt$ and $rr$
components of the Einstein equations,
yielding for the metric functions
\begin{equation}
\mu'= \frac{1}{2} N x^2 \phi'^2 + e^{2 \gamma \phi }
\left[ N (K'^2 +H'^2)
  + \frac{1}{8 x^2} \left(
     \left(K^2+H^2-4 \right)^2 +12 K^2 H^2 \right)\right]
\ , \label{eqmu} \end{equation}
\begin{equation}
 A' =  \frac{2}{x} \left[\frac{1}{2} x^2 \phi'^2
                 + e^{2 \gamma \phi } (K'^2 +H'^2) \right] A
\ . \label{eqa} \end{equation}
Here we have introduced the dimensionless coordinate
$x=(g/\sqrt{4\pi G}) r$
(the prime indicates the derivative
with respect to $x$),
the dimensionless mass function
\begin{equation}
\mu=\frac{g}{\sqrt{4\pi G}} m = \frac{g m_{\rm Pl}}{\sqrt{4\pi}} m
\  , \end{equation}
the dimensionless dilaton field $\phi = \sqrt{4\pi G} \Phi$,
and the dimensionless coupling constant
$\gamma =\kappa/\sqrt{4\pi G}$.
The choice $\gamma=1$ corresponds to string theory,
while $4+n$ dimensional Kaluza-Klein theory
has $\gamma^2=(2+n)/n$ \cite{gib1}.

For the matter field functions we obtain the equations
\begin{equation}
(e^{2 \gamma \phi } ANK')'
= e^{2 \gamma \phi } \frac{1}{4 x^2} A K \left( K^2+7H^2-4 \right)
\ , \end{equation}
\begin{equation}
(e^{2 \gamma \phi } ANH')'
= e^{2 \gamma \phi } \frac{1}{4 x^2} A H \left( H^2+7K^2-4 \right)
\ , \end{equation}
\begin{equation}
(AN x^2 \phi')' = 2 \gamma A e^{2 \gamma \phi }
\left[ N (K'^2 +H'^2)
  + \frac{1}{8 x^2} \left(
     \left(K^2+H^2-4 \right)^2 +12 K^2 H^2 \right)\right]
\ . \label{eqdil} \end{equation}
With help of eq.~(\ref{eqa}) the metric function $A$
can be eliminated from the matter field equations.
Note, that the equations are symmetric with respect to
an interchange of the functions $K(x)$ and $H(x)$,
and to the transformations $K(x) \rightarrow -K(x)$,
and $H(x) \rightarrow -H(x)$,
yielding degenerate solutions.

Let us now derive relations between the dilaton field
and the metric functions.
We first note, that the terms containing gauge field functions
in eq.~(\ref{eqmu})
for the metric function $\mu$
and in eq.~(\ref{eqdil})
for the dilaton function $\phi$ agree,
allowing us to replace the gauge field term
in the dilaton field equation.
We further note, that the expression for the curvature scalar,
obtained from the contracted Einstein equations,
does not involve the gauge field,
\begin{equation}
\frac{1}{2} A r^2 R = A N x^2 \phi'^2
\ . \end{equation}
With help of this relation we then find
for the dilaton field the equation
\begin{equation}
(AN x^2 \phi')' = \frac{1}{2} \gamma \left( x^2 (2 A' N + A N') \right)'
\ , \label{rel1} \end{equation}
or, after integration,
\begin{equation}
 \phi' = \frac{1}{2} \gamma \left( \ln ( A^2 N) \right)' + \frac{C}{A N x^2}
\ , \label{rel2} \end{equation}
where $C$ is an integration constant.
Note, that these relations hold in general for static,
spherically symmetric solutions with magnetic gauge fields
(and analogous relations for electric gauge fields).
We return to the relations (\ref{rel1}) and (\ref{rel2})
in the next sections after fixing the boundary conditions.

As in Einstein-Yang-Mills theory,
comparison of the equations of the SO(3) embedding
and those of the SU(2) embedding \cite{don,lav2,maeda,bizon3,neill}
shows, that to each SU(2) solution
there corresponds a ``scaled SU(2)'' solution
of the SO(3) system
with precisely double the mass of its SU(2) counterpart.
(With $x = 2 \tilde x$, $\mu(x) = 2 \tilde \mu(\tilde x)$,
$K(x) = 2 w(\tilde x)$, $H(x)=0$
and $\phi(x) = \tilde \phi(\tilde x)$
the functions $\tilde \mu$, $w$ and $\tilde \phi$
satisfy the SU(2) equations with coordinate $\tilde x$.)

\section{Regular Solutions}

Let us first consider the regular solutions
of SU(3) Einstein-Yang-Mills-dilaton theory.
Requiring asymptotically flat solutions implies
that the metric functions $A$ and $\mu$ both
approach a constant at infinity.
We here adopt
\begin{equation}
A(\infty)=1
\ , \end{equation}
thus fixing the time coordinate.
For magnetically neutral solutions
the gauge field functions approach a vacuum configuration
\begin{equation}
K(\infty)=\pm 2 \ , \quad H(\infty)=0
\ , \label{bc1} \end{equation}
\begin{equation}
K(\infty)=0 \ , \quad H(\infty)=\pm 2
\ . \label{bc2} \end{equation}
For the dilaton field we choose \cite{don,lav2,bizon3,neill}
\begin{equation}
\phi(\infty) = 0
\ . \label{bc5} \end{equation}

At the origin regularity of the solutions requires
\begin{equation}
\mu(0)=0
\ , \end{equation}
the gauge field functions must satisfy
\begin{equation}
K(0)=\pm 2 \ , \quad H(0)=0
\ , \label{bc3} \end{equation}
\begin{equation}
K(0)=0 \ , \quad H(0)=\pm 2
\ , \label{bc4} \end{equation}
while the dilaton field satisfies
\begin{equation}
\phi'(0) = 0
\ . \label{bc6} \end{equation}
As in Einstein-Yang-Mills theory \cite{kks}
it is sufficient to consider solutions
with $K(0)=2$ and $H(0)=0$.

We now return to the relations
between the metric and the dilaton field,
eqs.~(\ref{rel1}) and (\ref{rel2}).
Defining the dilaton charge $D$ via
\begin{equation}
\phi(x) \stackrel {x\rightarrow \infty} {\longrightarrow}
-\frac{D}{x}
\ , \label{bc7} \end{equation}
eq.~(\ref{rel1}) connects the dilaton charge and the mass
of the solution. By integrating
eq.~(\ref{rel1}) from zero to infinity
we obtain the relation
\begin{equation}
D = \gamma \mu(\infty)
\ . \label{res1} \end{equation}
Consequently the integration constant in eq.~(\ref{rel2})
vanishes, $C=D-\gamma \mu(\infty)$.
Integrating eq.~(\ref{rel2})
from $x$ to infinity then gives the relation
\begin{equation}
\phi(x) = \frac{1}{2} \gamma \ln( A^2 N)
   =\frac{1}{2} \gamma \ln(-g_{tt})
\ . \label{res2} \end{equation}
These results (\ref{res1}) and (\ref{res2})
generalize the relations obtained previously
for SU(2) and $\gamma=1$ \cite{don,bizon3,neill,gal}\footnote{
Ref.~\cite{bizon3} contains
a scaled version of relation (\ref{res1}) for general $\gamma$.}.
They are valid for general
static, spherically symmetric, magnetic gauge fields.

\begin{figure}
\centering
\epsfysize=11cm
\mbox{\epsffile{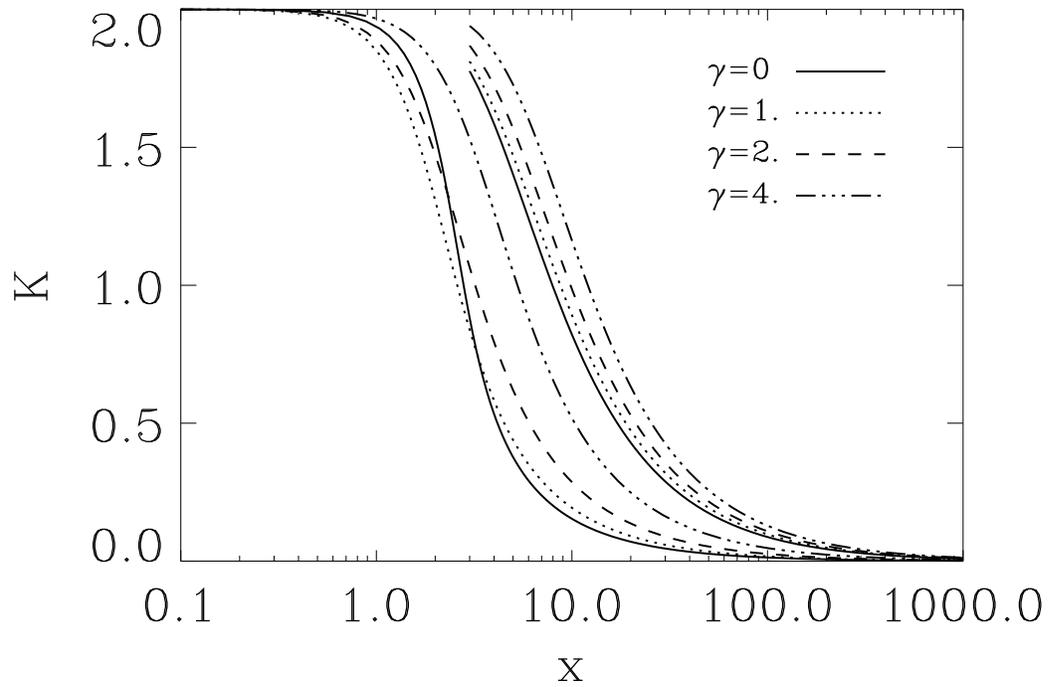}}
\caption{\label{k}
The function $K(x)$
is shown for the regular solution and for the
black hole solution with horizon $x_{\rm H}=3$
for the dilaton coupling constants $\gamma=0$ (solid),
$\gamma=1$ (dotted), $\gamma=2$ (dashed) and
$\gamma=4$ (tripledot-dashed)
as a function of $x$.}
\end{figure}

\begin{figure}
\centering
\epsfysize=11cm
\mbox{\epsffile{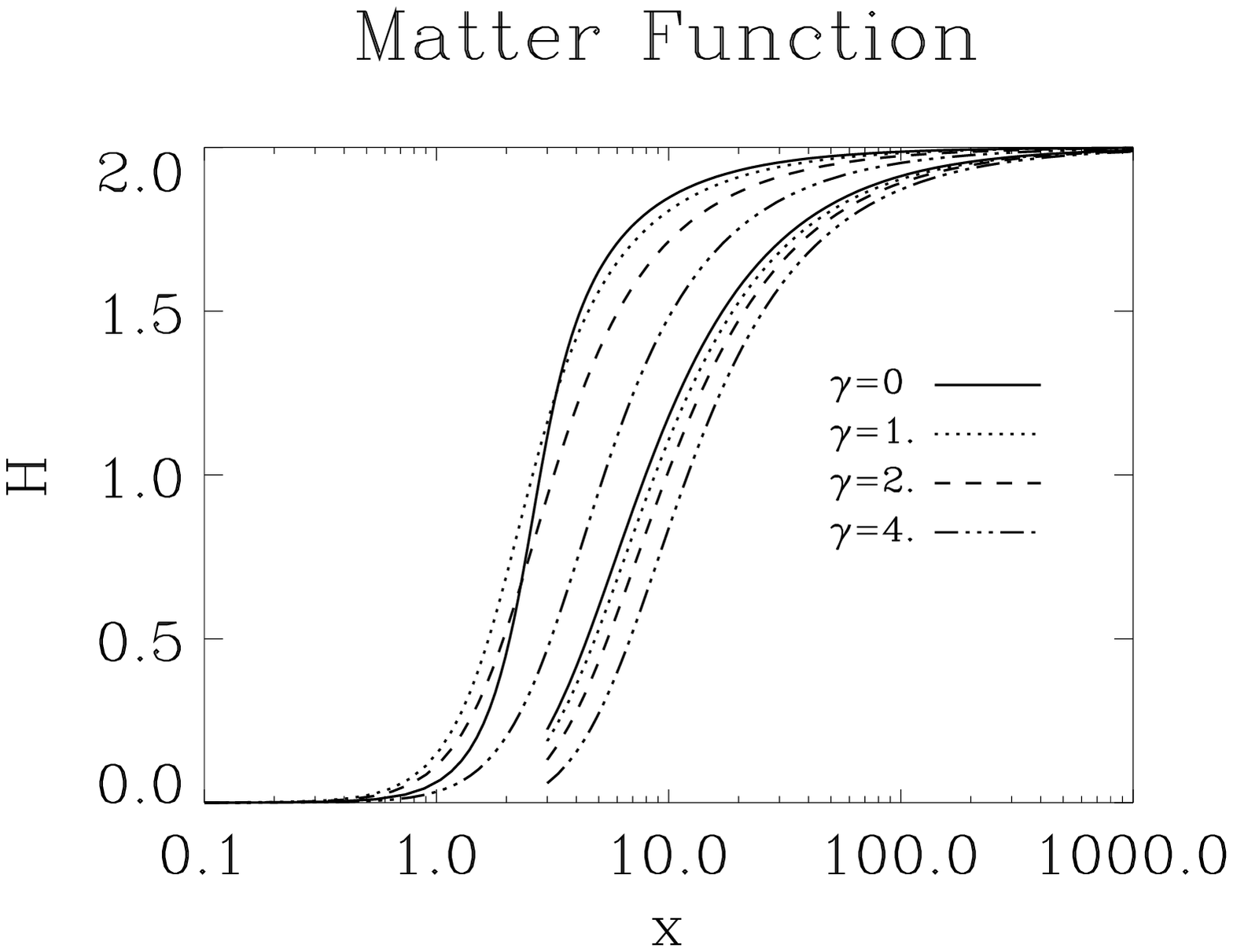}}
\caption{\label{h}
The function $H(x)$
is shown for the regular solution and for the
black hole solution with horizon $x_{\rm H}=3$
for the dilaton coupling constants $\gamma=0$ (solid),
$\gamma=1$ (dotted), $\gamma=2$ (dashed) and
$\gamma=4$ (tripledot-dashed)
as a function of $x$.}
\end{figure}

\begin{figure}
\centering
\epsfysize=11cm
\mbox{\epsffile{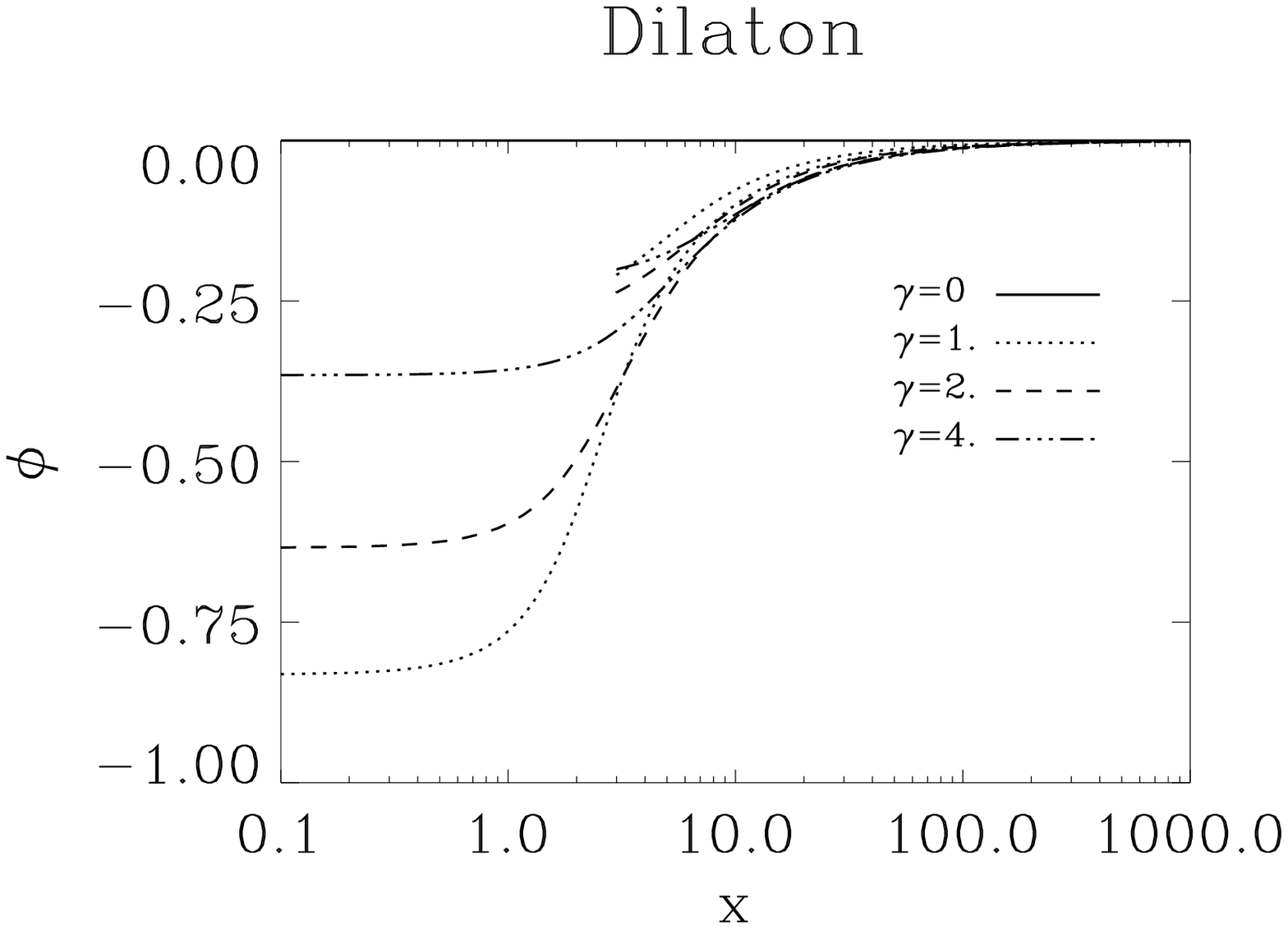}}
\caption{\label{p}
The function $\phi(x)$
is shown for the regular solution and for the
black hole solution with horizon $x_{\rm H}=3$
for the dilaton coupling constants $\gamma=0$ (solid),
$\gamma=1$ (dotted), $\gamma=2$ (dashed) and
$\gamma=4$ (tripledot-dashed)
as a function of $x$.}
\end{figure}

\begin{figure}
\centering
\epsfysize=11cm
\mbox{\epsffile{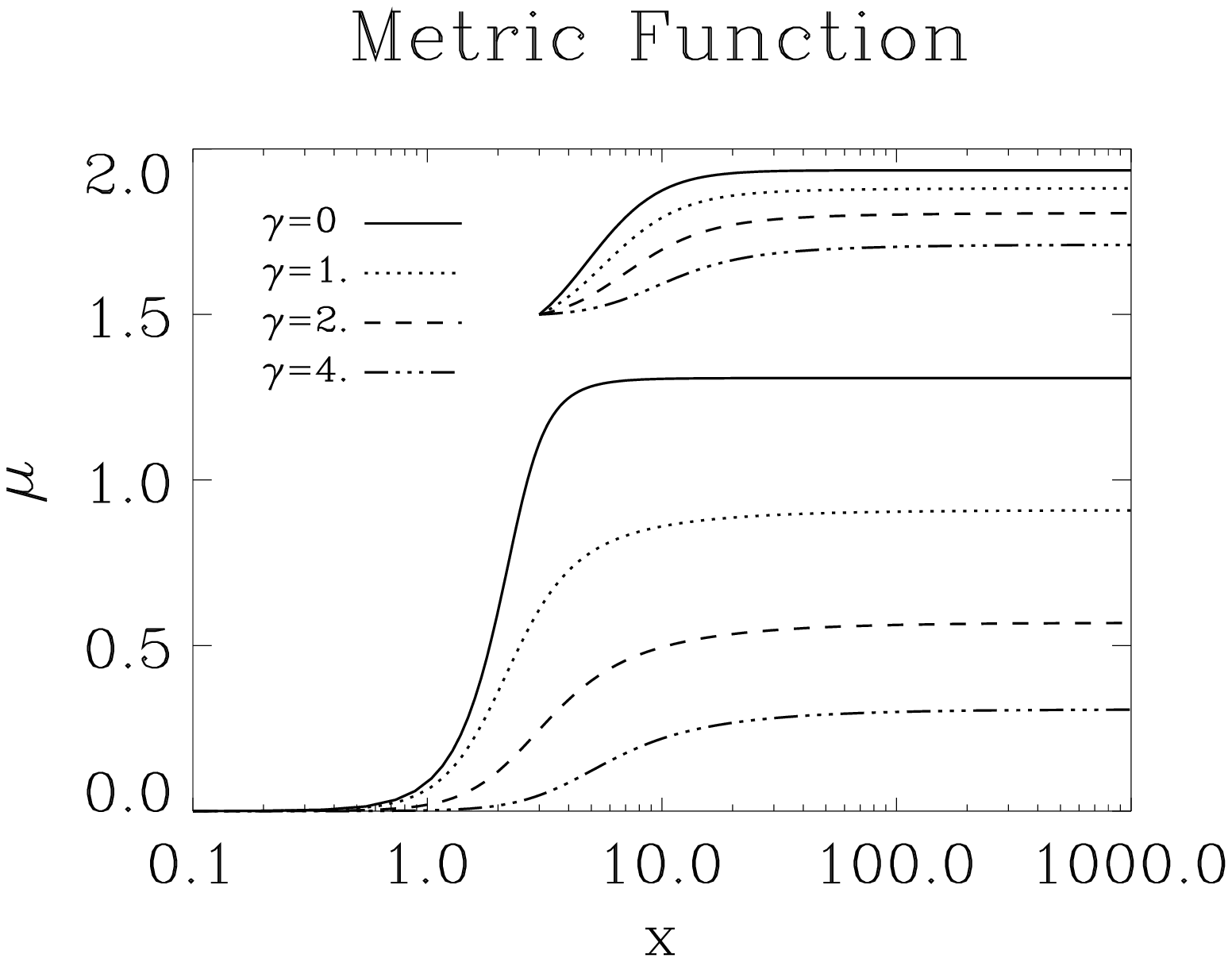}}
\caption{\label{mu}
The function $\mu(x)$
is shown for the regular solution and for the
black hole solution with horizon $x_{\rm H}=3$
for the dilaton coupling constants $\gamma=0$ (solid),
$\gamma=1$ (dotted), $\gamma=2$ (dashed) and
$\gamma=4$ (tripledot-dashed),
as a function of $x$.}
\end{figure}

In the following we present numerical results
for the regular solutions of the SO(3) embedding.
In Figs.~\ref{k}-\ref{mu} we show the lowest SO(3) solution
for $\gamma=0$, 1, 2 and 4.
The excited solutions will be shown elsewhere \cite{we}.
In the limit of vanishing coupling constant $\gamma$
the solutions approach smoothly the corresponding
Einstein-Yang-Mills solutions \cite{kks}.
In the limit $\gamma \rightarrow \infty$,
on the other hand, the solutions approach those of
SU(3) Yang-Mills-dilaton theory in flat space.

\begin{figure}
\centering
\epsfysize=11cm
\mbox{\epsffile{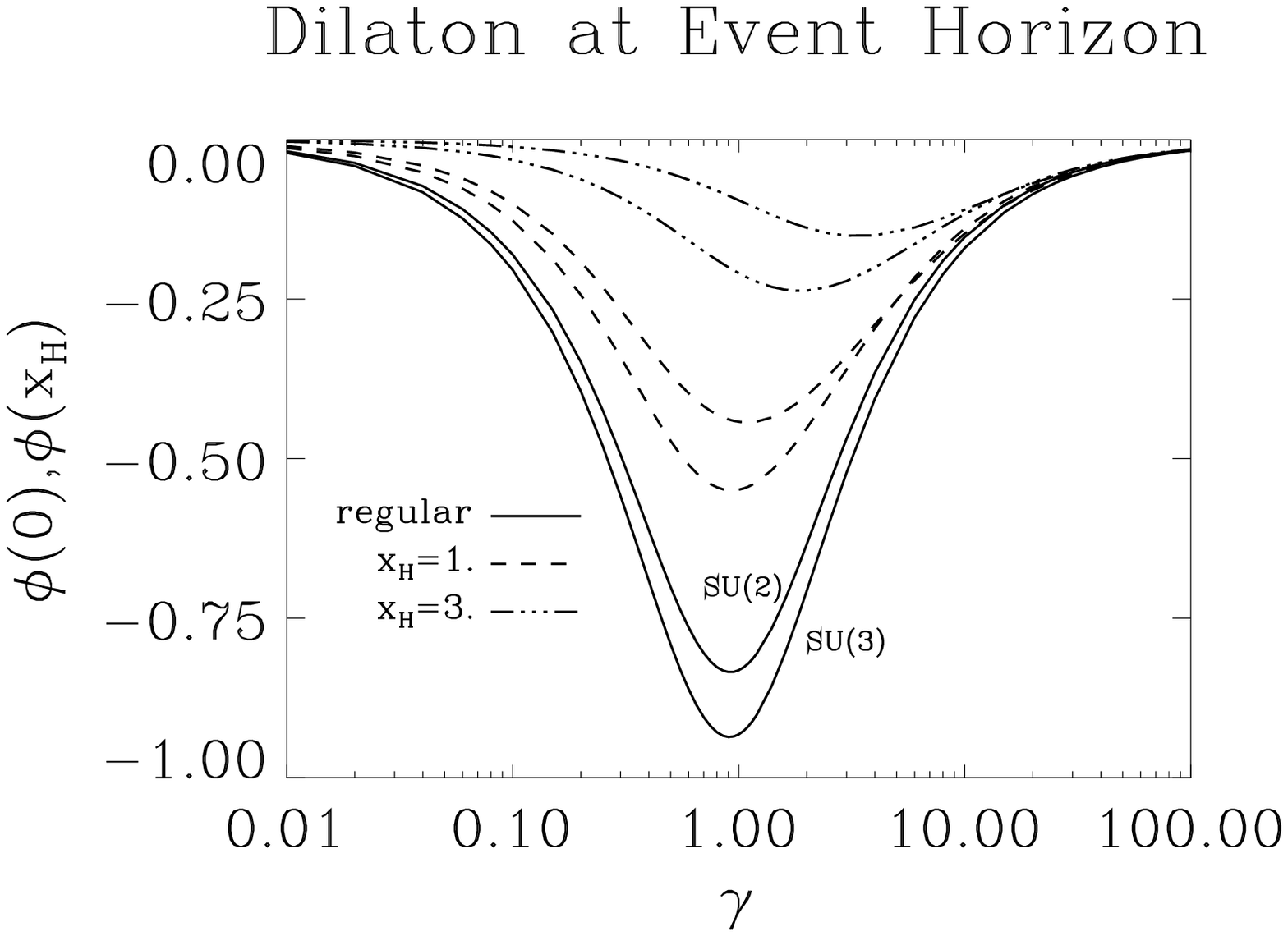}}
\caption{\label{phx}
The dilaton field at the origin, $\phi(0)$,
for the regular solution (solid)
and the dilaton field at the horizon, $\phi(x_{\rm H})$,
for the black hole solutions with horizons
$x_{\rm H}=1$ (dashed) and $x_{\rm H}=3$ (tripledot-dashed)
are shown as a function of the dilaton coupling constant $\gamma$.
The lower curves correspond to SO(3) solutions, while the upper
curves correspond to SU(2) solutions.}
\end{figure}

In Fig.~\ref{phx}
we show the value of the dilaton field at the origin
as a function of the coupling constant $\gamma$,
for the lowest SO(3) and SU(2) solutions.
While in the limits $\gamma \rightarrow 0$ and
$\gamma \rightarrow \infty$ the value of the dilaton field
at the origin approaches zero,
a minimum occurs at $\gamma=0.92$
for the SO(3) solutions,
close to the minimum of the SU(2) solutions at $\gamma=0.91$.
Remember, that the dilaton function is related to
the metric via eq.~(\ref{res2}).

To discuss the sequences of solutions,
we adopt the classification of
the solutions with respect to
the number of nodes $(n_1,n_2)$ of the functions $(u_1,u_2)$
\cite{kuenzle,kks},
\begin{equation}
u_1(x)= \frac{K(x)+H(x)}{2} \ , \ \ \
u_2(x)= \frac{K(x)-H(x)}{2}
\ , \end{equation}
and the total number of nodes $n$.
($n=n_1+n_2$ for SO(3) solutions.)
The lowest SO(3) solution
has $n=1$ and node structure $(0,1)$,
while the lowest scaled SU(2) solution
has $n=2$ and node structure $(1,1)$.

As observed in \cite{lav2,bizon3},
in the limit $n \rightarrow \infty$
the sequence of regular SU(2) solutions for a given $\gamma$
tends to the ``extremal'' Einstein-Maxwell-dilaton black hole
\cite{gib1,str} with the same $\gamma$
and with magnetic charge $P=1$.
``Extremal'' Einstein-Maxwell-dilaton black hole solutions
satisfy the relation between mass and magnetic charge,
$\mu=P/\sqrt{1+\gamma^2}$.
The sequences of scaled SU(2) solutions,
having twice the mass and dilaton charge
of the corresponding SU(2) solutions, therefore approach
the ``extremal'' Einstein-Maxwell-dilaton black holes
with twice the magnetic charge, $P=2$.
Consequently, the limiting mass
of this sequence of solutions with node structure $(n,n)$
is given by $\mu=2/\sqrt{1+\gamma^2}$.

In Table~1 the dimensionless mass $\mu(\infty)$
of the lowest SO(3) solutions with four or less nodes is shown
for the values of the dilaton coupling constant
$\gamma=0$, 1, 3, and 10.
Also shown is the limiting case $\gamma=\infty$.
This flat space limit is obtained by scaling
the coordinate, the mass and the dilaton field according to
$x= \bar x \gamma$, $\mu= \bar \mu / \gamma$, and
$\phi= \bar \phi / \gamma$.
The ADM mass of the solutions is given by
$m_{\rm ADM}= \mu(\infty) {\sqrt{4 \pi} m_{\rm Pl}}/{g}$.

Let us now inspect the sequence of solutions
with node structure $(0,n)$,
including the lowest SO(3) solution.
The masses $\mu(\infty)$ of these solutions,
shown in Table~1 for $n=1-4$,
are well approximated by the formula
\begin{equation}
\mu(\infty,\gamma,n)=\frac{\sqrt{3}-\frac{\pi}{2} e^{-\frac{4}{3}n}}
            {\sqrt{1+\gamma^2}}
\ . \label{eng} \end{equation}
They approach for large $n$
the (extrapolated) limiting value $\sqrt{3/(1+\gamma^2)}$,
shown for comparison.
This indicates, that in the limit $n \rightarrow \infty$
the regular SO(3) solutions with node structure $(0,n)$ tend
to the ``extremal'' Einstein-Maxwell-dilaton black holes
\cite{gib1,str} with magnetic charge $P=\sqrt{3}$ \cite{we}.

The order of the solutions with respect to their mass
does not depend on the dilaton coupling constant.
For a given total number of nodes,
the $(0,n)$ solutions are the lowest solutions \cite{kks},
as seen in Table~1,
and their limiting mass is lower than
the limiting mass of the $(n,n)$ solutions.
Further details will be given elsewhere \cite{we}.
The limiting behaviour
of solutions with general node structure $(n_1,n_2)$
is still open, since excited solutions
are difficult to obtain.

The regular SU(2) Einstein-Yang-Mills-dilaton solutions
are unstable \cite{lav2,bizon3},
so are the regular SO(3) Einstein-Yang-Mills solutions,
obtained in the limit $\gamma \rightarrow 0$
\cite{kks,strau3}.
By continuity, we conclude, that the
regular SO(3) Einstein-Yang-Mills-dilaton solutions
are unstable, too.

\section{Black Hole Solutions}

We now turn to the black hole solutions of
SU(3) Einstein-Yang-Mills-dilaton theory.
Imposing again the condition of asymptotic flatness,
the black hole solutions considered here satisfy the same
boundary conditions at infinity
as the regular solutions.
The existence of a regular event horizon at $x_{\rm H}$
requires
\begin{equation}
\mu(x_{\rm H})= \frac{x_{\rm H}}{2}
\ , \end{equation}
and $A(x_{\rm H}) < \infty $,
and the matter functions at the horizon $x_{\rm H}$ must satisfy
\begin{eqnarray}
 N'K'= \frac{1}{4 x^2} K \left( K^2+7H^2-4 \right)
\ , \\
 N'H'= \frac{1}{4 x^2} H \left( H^2+7K^2-4 \right)
\ , \\
 N'\phi'= \gamma e^{2 \gamma \phi }
\left[
   \frac{1}{4 x^4} \left(
     \left(K^2+H^2-4 \right)^2 +12 K^2 H^2 \right)\right]
\ .
\end{eqnarray}
Note that a restricted subset of Einstein-Yang-Mills-dilaton
black hole solutions with $K \equiv H$ (and thus different
boundary conditions),
corresponding to magnetically charged black holes,
was obtained previously \cite{gal}.
Unlike the magnetically neutral black hole solutions
considered here, however,
these solutions have no regular limit \cite{volkov3}.

Let us introduce the dimensionless Hawking temperature $T$
for the metric (\ref{metric}) \cite{don,maeda}
\begin{equation}
T = T_{\rm S} A (1-2 \mu')|_{x_{\rm H}}
  = T_{\rm S} x_{\rm H} A N' |_{x_{\rm H}}
\ , \label{thaw} \end{equation}
where $T_{\rm S} = (4 \pi x_{\rm H})^{-1}$
is the Hawking temperature of the Schwarzschild black hole.
Let us now consider the relations (\ref{rel1}) and (\ref{rel2})
for black holes.
Integrating eq.~(\ref{rel1})
from the horizon to infinity yields the relation
\begin{equation}
D = \gamma \left( \mu(\infty) - \frac{1}{2} x_{\rm H}^2 A N' |_{x_{\rm H}}
  \right)
  = \gamma \mu(\infty) \left( 1 - \frac{\mu_{\rm S}}{\mu(\infty)}
   \frac{T}{T_{\rm S}}  \right)
\ , \label{res3} \end{equation}
where $\mu_{\rm S}= x_{\rm H}/2$
is the mass of the Schwarzschild black hole.
An equivalent expression was obtained for SU(2)
black holes for $\gamma=1$ \cite{neill}.
Again relation (\ref{res3}) holds for general static,
spherically symmetric, magnetic gauge fields,
i.~e.~beside the magnetically neutral SU(2) and SO(3) black holes
considered here,
it also holds for the magnetically charged Einstein-Maxwell-dilaton
black holes \cite{gib1,str} and
SU(3) Einstein-Yang-Mills dilaton black holes
\cite{gal}.
Unlike the regular case, the integration constant
$C=D-\gamma \mu(\infty)$ here in general
does not vanish\footnote{
An exception are the ``extremal'' magnetically charged
Einstein-Maxwell-dilaton black holes \cite{gib1,str}.},
and integration of eq.~(\ref{rel2})
does not yield a simple expression.

The SO(3) Einstein-Yang-Mills-dilaton black hole solutions
are similar to the Einstein-Yang-Mills black hole solutions \cite{kks}
and approach these smoothly in the limit $\gamma \rightarrow 0$.
As examples we show the radial functions for the lowest
SO(3) black hole solutions with horizon $x_{\rm H}=3$
and $\gamma=0$, 1, 2 and 4 in Figs.~(\ref{k})-(\ref{mu}).
In Fig.~(\ref{phx}) we show the dilaton field at
the horizon. With increasing horizon
the minimum of $\phi(x_{\rm H})$ shifts to larger values of $\gamma$.
Further details of these solutions
and the excited SO(3) Einstein-Yang-Mills-dilaton black holes
will be shown elsewhere \cite{we}.

The families of SO(3) Einstein-Yang-Mills-dilaton black hole solutions
change continuously as a function of the horizon $x_{\rm H}$.
For vanishing horizon,
$x_{\rm H} \rightarrow 0$, the black hole solutions
approach the regular solutions.
With increasing horizon $x_{\rm H}$ the black hole solutions
keep their identity in terms of the boundary conditions
and the node structure.
Only the excited solutions with node structure $(n,n)$,
having a slightly higher mass
than the corresponding scaled SU(2) solutions,
merge at some critical values of the horizon
into the scaled SU(2) solutions \cite{kks}.
These critical values now depend on the dilaton coupling constant.
The critical value of the $(1,1)$ solution, for instance,
first decreases for small values of $\gamma$,
reaches a minimum at about $\gamma=1$, and then increases.
Note that the order of the solutions (Table~2) changes
from the order of the regular solutions
as the horizon increases.

In Table~2 we present properties of black hole solutions
with horizon $x_{\rm H}=1$,
emerging from the lowest regular solutions.
The table gives the mass $\mu(\infty)$,
the dilaton charge $D$, and the ratio of the Hawking temperature
to the Schwarzschild Hawking temperature,
for the dilaton coupling constants
$\gamma=1$ and 10.
As for the regular solutions,
we observe two different sets of limiting values
for the sequences of black hole solutions with node structure
$(0,n)$ and $(n,n)$.

Though not observed previously,
the generalization of the limiting behaviour
from the regular solutions to the black hole solutions
is straightforward.
The sequences of black hole solutions
tend to Einstein-Maxwell-dilaton black hole solutions
with the same $\gamma$ and the same horizon,
and with magnetic charge $P=1$ for the SU(2) solutions,
and magnetic charges $P=\sqrt{3}$ and $P=2$
for the $(0,n)$ and $(n,n)$ SO(3) solutions, respectively.
The limiting values are also shown in Table 2\footnote{
The properties of the limiting charged SU(3) solutions \cite{gal}
are obtained with $P=1$, not $P=\sqrt{3}/2$.}.
For $\gamma=1$ simple formulae hold for mass, dilaton charge
and Hawking temperature of the limiting black hole solutions
\cite{gib1,str}.
Denoting $r_+=\sqrt{x_{\rm H}^2+2P^2}$
one obtains $\mu(\infty)=r_+/2$, $D=P^2/r_+$
and $T/T_{\rm S}=x_{\rm H}/r_+$.

\begin{figure}
\centering
\epsfysize=11cm
\mbox{\epsffile{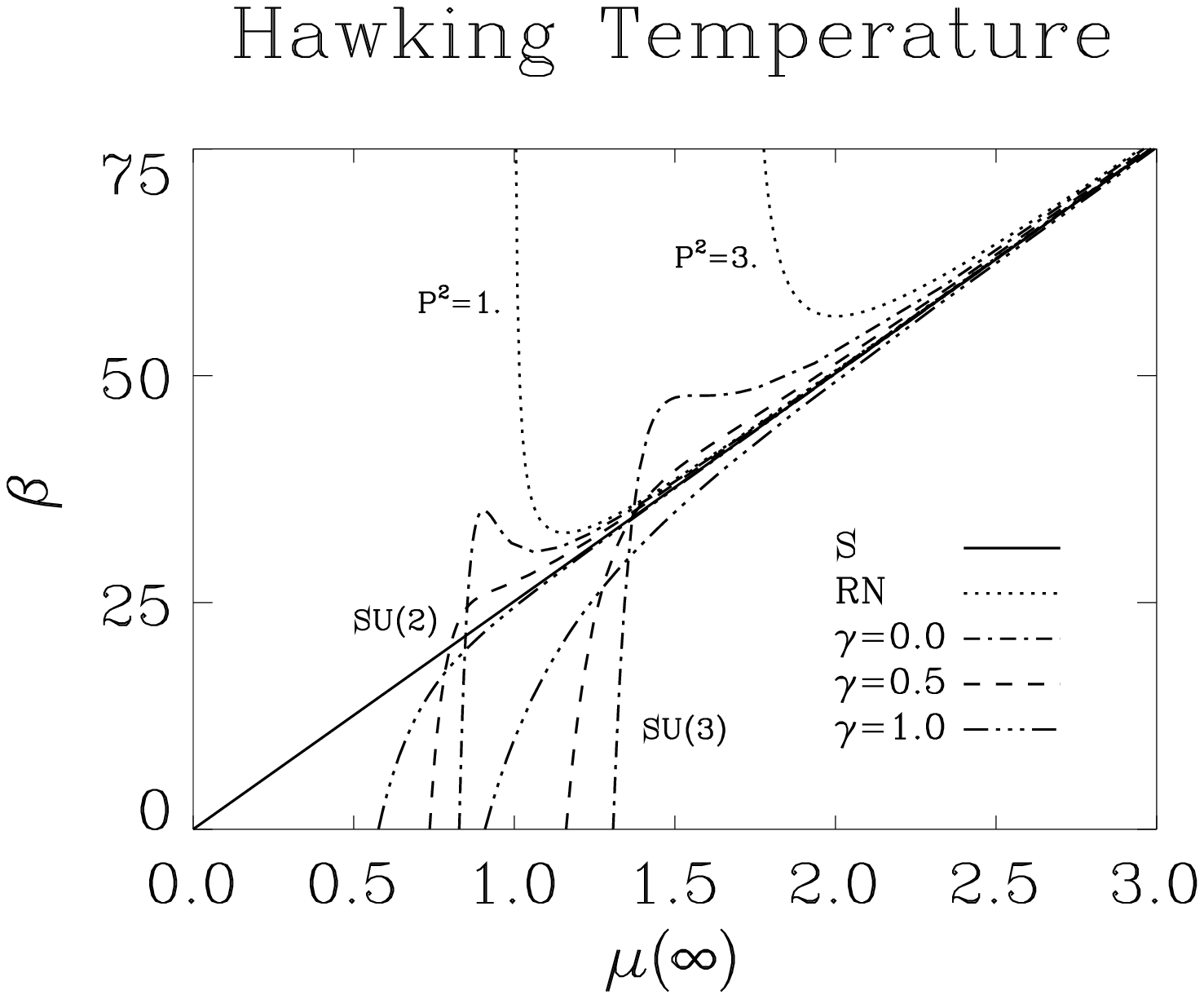}}
\caption{\label{beta}
The inverse of the dimensionless Hawking temperature $\beta=T^{-1}$
is shown as a function of the dimensionless
black hole mass $\mu(\infty)$
for the lowest SU(2) and SO(3) solutions
and for dilaton coupling constants $\gamma=0$ (dot-dashed),
$\gamma=0.5$ (dashed), and $\gamma=1$ (tripledot-dashed).
Also shown is the inverse temperature for
the Schwarzschild (solid)
and Reissner-Nordstr\o m black holes (dotted)
with magnetic charge P.}
\end{figure}

Let us now turn to the thermodynamic properties
of the black hole solutions.
In Fig.~\ref{beta} we show the dimensionless inverse Hawking temperature
of the lowest black hole solutions for SU(2) and SO(3)
as a function of their mass,
for the dilaton coupling constants $\gamma=0$, 0.5 and 1.
Also shown are the corresponding Schwarzschild
and Reissner-Nordstr\o m inverse Hawking temperatures.
As in the SU(2) case \cite{maeda}\footnote{
The coupling of ref.~\cite{maeda}
is related to ours via $\kappa = \sqrt{2}\gamma$.}, we observe,
depending on the dilaton coupling constant,
phase transitions.
For vanishing $\gamma$, there
are two critical values
of the mass $\mu_1=1.561$ and $\mu_2=1.588$ for SO(3)
and $\mu_1=0.905$ and $\mu_2=1.061$ for SU(2) \cite{maeda}.
The critical behaviour disappears beyond $\gamma=0.0293$
for SO(3) and $\gamma=0.38$ for SU(2).
The thermodynamical properties of the excited solutions
tend with increasing $n$ to
those of the corresponding Einstein-Maxwell-dilaton black holes
with magnetic charges $P=1$ for SU(2)
and $P=\sqrt{3}$ and $P=2$
for the $(0,n)$ and $(n,n)$ SO(3) sequences, respectively,
while for $\gamma=0$ they tend to those of
the corresponding Reissner-Nordstr\o m black holes
\cite{we}.
Further details will be given elsewhere \cite{we}.

The SU(2) Einstein-Yang-Mills-dilaton black holes
are unstable \cite{lav2},
so are the SO(3) Einstein-Yang-Mills black holes
obtained in the limit $\gamma \rightarrow 0$
\cite{kks,strau3}.
Again we conclude by continuity, that the
SO(3) Einstein-Yang-Mills-dilaton black hole solutions
are also unstable.

\section{Conclusion}

Classical solutions of SU(3) Einstein-Yang-Mills-dilaton theory
are very similar to those of SU(3) Einstein-Yang-Mills theory
\cite{kks}.
There are sequences of regular and black hole solutions
based on static, spherically symmetric, magnetic ans\"atze
for the SO(3) embedding
as well as the previously studied
SU(2) embedding \cite{don,lav2,maeda,bizon3,neill}.

For the regular solutions two relations between
metric and dilaton field hold,
$D=\gamma \mu(\infty)$
and $\phi(x)=\gamma \ln(\sqrt{-g_{tt}})$,
whereas the black hole solutions satisfy the relation between
mass and dilaton charge,
$D=\gamma (\mu(\infty)-2\pi x_{\rm H}^2 T)$,
where $T$ is the Hawking temperature.
These relations holds in general for static,
spherically symmetric solutions
with magnetic gauge fields.

The SO(3) solutions can be labeled
according to the node structure
of the gauge field functions by the integers $(n_1,n_2)$
and the total number of nodes $n$.
The lowest solutions have
node structure $(0,1)$ and $(0,2)$,
being of type $(0,n)$.
The next two solutions both have node structure $(1,1)$,
corresponding to
the lowest scaled SU(2) solution and an excitation.

For large $n$ the solutions approach limiting solutions.
In the limit $n \rightarrow \infty$,
mass and dilaton charge of the regular excited SO(3) solutions
with node structure $(0,n)$ and $(n,n)$
tend to those of the ``extremal'' Einstein-Maxwell-dilaton solutions
with magnetic charge $P=\sqrt{3}$ and $P=2$, respectively,
while mass and dilaton charge
of the regular excited SU(2) solutions
tend to those of the ``extremal'' Einstein-Maxwell-dilaton
solutions with magnetic charge $P=1$.

For the black hole solutions the limiting behaviour is analogous.
In the limit $n \rightarrow \infty$
mass, dilaton charge and Hawking temperature
of a sequence of excited black hole solutions with a given horizon
tend to those of the corresponding Einstein-Maxwell-dilaton solution
with the same horizon.
(The limiting behaviour of the metric and matter functions
will be discussed elsewhere \cite{we}.)

The lowest SO(3) Einstein-Yang-Mills black holes
have similar thermodynamic properties as their
SU(2) counterparts \cite{maeda}.
For small $\gamma$, there are two critical masses
associated with two phase transitions,
where the specific heat changes sign.
The thermodynamic properties of the
excited black hole solutions tend to those
of the corresponding Einstein-Maxwell-dilaton solutions
with magnetic charges $P=1$, $P=\sqrt{3}$ and $P=2$
for the SU(2) and SO(3) $(0,n)$ and $(n,n)$ solutions, respectively.

We finally remark that
the SO(3) Einstein-Yang-Mills-dilaton black hole solutions
constructed here constitute counterexamples
to the ``no-hair conjecture'',
since SU(3) Einstein-Yang-Mills-dilaton theory also contains
Schwarz\-schild black holes.
However, only the Schwarz\-schild black holes are stable.

{\sl Acknowledgement}

We dedicate this work to Larry Wilets
on the occasion of his retirement.
We gratefully acknowledge discussions with M. Volkov.

\newpage
\begin{table}
\begin{center}
\begin{tabular}{|c|ccccccccc|} \hline
\multicolumn{1}{|c|}{\# }&
\multicolumn{2}{c}{nodes} &        &
\multicolumn{4}{c}{$\mu(\infty)$} & &
\multicolumn{1}{c|}{$\bar{\mu}(\infty)$}   \\
\cline{2-3} \cline{5-8} \cline{10-10}
    & $u_1$        & $u_2$      & $\gamma$ = &
$0.$      &   $1.$     &    $3.$      &   $10.$     &  &   $\infty$  \\
\hline
$1$& $0$          & $1$        &            &
$1.30778$ &   $0.90853$ &   $0.40083$ &   $0.12575$ &  &   $1.26319$ \\
$2$& $0$          & $2$        &            &
$1.62261$ &   $1.14153$ &   $0.50845$ &   $0.15984$ &  &   $1.60603$ \\
$3^{\ast} $& $1$          & $1$        &            &
$1.65729$ &   $1.15397$ &   $0.50989$ &   $0.16001$ &  &   $1.60753$ \\
$4        $& $1$          & $1$        &            &
$1.69538$ &   $1.18798$ &   $0.52760$ &   $0.16576$ &  &   $1.66536$ \\
$5$& $0$          & $3$        &            &
$1.70474$ &   $1.20383$ &   $0.53779$ &   $0.16918$ &  &   $1.69998$ \\
$6$& $0$          & $4$        &            &
$1.72531$ &   $1.21958$ &   $0.54526$ &   $0.17156$ &  &   $1.72393$ \\
$l_1$& $0$          & $\infty$   &            &
$1.73205$ &   $1.22474$ &   $0.54772$ &   $0.17235$ &  &   $1.73205$ \\
$7        $& $1$          & $2$        &            &
$1.87547$ &   $1.32130$ &   $0.58916$ &   $0.18526$ &  &  $1.86143$  \\
$8        $& $1$          & $3$        &            &
$1.92840$ &   $1.36196$ &   $0.60850$ &   $0.19143$ &  &  $1.92350$  \\
$9^{\ast} $& $2$          & $2$        &            &
$1.94269$ &   $1.36967$ &   $0.61107$ &   $0.19217$ &  &  $1.93108$  \\
$10       $& $2$          & $2$        &            &
$1.94707$ &   $1.37417$ &   $0.61363$ &   $0.19302$ &  &  $1.93963$  \\
$l_2      $& $\infty$     & $\infty$   &            &
$2.00000$ &   $1.41421$ &   $0.63246$ &   $0.19901$ & &   $2.00000$  \\
\hline
\end{tabular}
\end{center}
\vspace{1.cm}

{\bf Table 1}\\
The dimensionless mass $\mu(\infty)$ of the lowest regular
SO(3) solutions with four or less nodes
for dilaton coupling constants $\gamma=0$, 1, 3, 10,
and the scaled mass $\bar\mu(\infty)$, obtained
in the flat space limit $\gamma \rightarrow \infty$.
The lines $l_1$ and $l_2$ give
the limiting values $\sqrt{3/(1+\gamma^2)}$
and $\sqrt{4/(1+\gamma^2)}$
of the $(0,n)$ and $(n,n)$ sequences, respectively.
The $\ast$ indicates the scaled SU(2) solutions.
The numbers \# refer to the order of these solutions
with respect to their mass.
\vspace{1.cm} \\
\end{table}
\begin{table}
\begin{center}
\begin{tabular}{|c|ccccccccc|} \hline
\multicolumn{1}{|c|} {\# }&         &
\multicolumn{2}{c}{$\mu(\infty)$} & &
\multicolumn{2}{c}{$D$} & &
\multicolumn{2}{c|}{$T/T_S$} \\
\cline{3-4} \cline{6-7} \cline{9-10}
    & $\gamma$ = &   $1.$     &   $10.$     &
      $\gamma$ = &   $1.$     &   $10.$     &
      $\gamma$ = &   $1.$     &   $10.$     \\ \hline
$1$       &             &
$1.16497$ &   $ .61858$ &            &
$.87027$  &   $1.25371$ &            &
$.58942$  &   $.98641$  \\
$2$       &             &
$1.30296$ &   $ .64749$ &            &
$1.08770$ &   $1.58722$ &            &
$.43054$  &   $ .97754$ \\
$3^{\ast}$&             &
$1.38135$ &   $ .65196$ &            &
$1.12231$ &   $1.59670$ &            &
$.51809$  &   $ .98458$ \\
$4$       &             &
$1.38437$ &   $ .65546$ &            &
$1.13889$ &   $1.64944$ &            &
$.49097$  &   $ .98104$ \\
$5$       &             &
$1.32140$ &   $ .65284$ &            &
$1.12979$ &   $1.66560$ &            &
$.38322$  &   $ .97255$ \\
$6$       &             &
$1.32279$ &   $ .65335$ &            &
$1.13364$ &   $1.67533$ &            &
$.37830$  &   $ .97163$ \\
$l_1$     &             &
$1.32288$ &   $ .65339$ &            &
$1.13389$ &   $1.67603$ &            &
$.37796$  &   $ .97157$ \\
$7$       &             &
$1.46227$ &   $ .67144$ &            &
$1.26493$ &   $1.83794$ &            &
$.39467$  &   $ .97529$ \\
$8$       &             &
$1.47209$ &   $ .67440$ &            &
$1.28839$ &   $1.88424$ &            &
$.36740$  &   $ .97195$ \\
$9^{\ast}$&             &
$1.49382$ &   $ .67739$ &            &
$1.31695$ &   $1.90744$ &            &
$.35375$  &   $ .97329$ \\
$l_2$     &             &
$1.50000$ &   $ .67936$ &            &
$1.33333$ &   $1.94262$ &            &
$.33333$  &   $ .97020$ \\
\hline
\end{tabular}
\end{center}
\vspace{1.cm}
{\bf Table 2}\\
The dimensionless mass $\mu(\infty)$, dilaton charge $D$ and
ratio of the Hawking temperature
to the Schwarzschild Hawking temperature $T/T_{\rm S}$
of the lowest black hole solutions with four or less nodes
for dilaton coupling constants $\gamma=1$ and 10
and horizon $x_{\rm H} = 1$.
The lines $l_1$ and $l_2$ give the limiting values
of the $(0,n)$ and $(n,n)$ sequences, respectively.
The $\ast$ indicates the scaled SU(2) solutions.
The numbers \#  correspond to those of the
regular solutions (obtained for $x_{\rm H} \rightarrow 0$).
Note that the black hole solution $\# 10$ with node structure $(2,2)$
does not exist for $0.4 < \gamma < 10.5$ and horizon $x_{\rm H} = 1$.
\vspace{1.cm} \\

\end{table}
\newpage

\newpage

\newpage

\newpage

\newpage

\begin{thebibliography}{000}
\bibitem{bm}
 R. Bartnik and J. McKinnon,
 Particlelike solutions of the Einstein-Yang-Mills equations,
 Phys. Rev. Lett. 61 (1988) 141.
\bibitem{strau1}
 N. Straumann and Z.~H. Zhou,
 Instability of the Bartnik-McKinnon solutions
 of the Einstein-Yang-Mills equations,
 Phys. Lett. B237 (1990) 353.
\bibitem{volkov1}
 D.~V. Gal'tsov and M.~S. Volkov,
 Sphalerons in Einstein-Yang-Mills theory,
 Phys. Lett. B273 (1991) 255.
\bibitem{lav1}
 G. Lavrelashvili and D. Maison,
 Static spherically symmetric solutions of a Yang-Mills field
 coupled to a dilaton,
 Phys. Lett. B295 (1992) 67.
\bibitem{bizon2}
 P. Bizon,
 Saddle-point solutions in Yang-Mills-dilaton theory,
 Phys. Rev. D47 (1993) 1656.
\bibitem{dhn}
 R.~F. Dashen, B. Hasslacher and A. Neveu,
 Nonperturbative methods and extended-hadron models
 in field theory. III.~Four-dimensional non-abelian models,
 Phys. Rev. D12 (1974) 4138.
\bibitem{bog}
 J. Boguta,
 Can nuclear interactions be long-ranged?
 Phys. Rev. Lett. 50 (1983) 148.
\bibitem{km}
 F.~R. Klinkhamer and N.~S. Manton,
 A saddle-point solution in the Weinberg-Salam theory,
 Phys. Rev. D30 (1984) 2212.
\bibitem{kb1}
 J. Kunz and Y. Brihaye,
 New sphalerons in the Weinberg-Salam theory,
 Phys. Lett. B216 (1989) 353.
\bibitem{yaffe}
 L.~G. Yaffe,
 Static solutions of SU(2)-Higgs theory,
 Phys. Rev. D40 (1989) 3463.
\bibitem{don}
 E.~E. Donets and D.~V. Gal'tsov,
 Stringy sphalerons and non-abelian black holes,
 Phys. Lett. B302 (1993) 411.
\bibitem{lav2}
 G. Lavrelashvili and D. Maison,
 Regular and black hole solutions of Einstein-Yang-Mills
 dilaton theory,
 Nucl. Phys. B410 (1993) 407.
\bibitem{maeda}
 T. Torii and K. Maeda,
 Black holes with non-Abelian hair and their thermodynamical
 properties,
 Phys. Rev. D48 (1993) 1643.
\bibitem{bizon3}
 P. Bizon,
 Saddle points of stringy action,
 Acta Physica Polonica B24 (1993) 1209.
\bibitem{neill}
 C.~M. O'Neill,
 Einstein-Yang-Mills theory with a massive dilaton and axion:
 String-inspired regular and black hole solutions,
 Phys. Rev. D50 (1994) 865.
\bibitem{volkov}
 M.~S. Volkov and D.~V. Galt'sov,
 Black holes in Einstein-Yang-Mills theory,
 Sov. J. Nucl. Phys. 51 (1990) 747.
\bibitem{bizon1}
 P. Bizon,
 Colored black holes,
 Phys. Rev. Lett. 64 (1990) 2844.
\bibitem{kuenzle1}
 H.~P. K\"unzle and A.~K.~M. Masoud-ul-Alam,
 Spherically symmetric static SU(2) Einstein-Yang-Mills fields,
 J. Math. Phys. 31 (1990) 928.
\bibitem{strau2}
 N. Straumann and Z.~H. Zhou,
 Instability of colored black hole solutions,
 Phys. Lett. B243 (1990) 33.
\bibitem{volkov5}
 M.~S. Volkov and D.~V. Gal'tsov,
 Odd-parity negative modes of Einstein-Yang-Mills
 black holes and sphalerons,
 Phys. Lett. B341 (1995) 279.
\bibitem{volkov4}
 M.~S. Volkov, O. Brodbeck, G. Lavrelashvili and N. Straumann,
 The number of sphaleron instabilities of the Bartnik-McKinnon solitons
 and nonabelian black holes,
 preprint ZU-TH-3-95, hep-th/9502045.
\bibitem{kuenzle}
 H.~P. K\"unzle,
 Analysis of the static spherically symmetric
 SU(n)-Einstein-Yang-Mills equations,
 Comm. Math. Phys. 162 (1994) 371.
\bibitem{kks}
 B. Kleihaus, J. Kunz and A. Sood,
 SU(3) Einstein-Yang-Mills sphalerons and black holes,
 Phys. Lett. B354 (1995) 240.
\bibitem{gib1}
 G.~W. Gibbons and K. Maeda,
 Black holes and membranes in higher-dimensional
 theories with dilaton fields,
 Nucl. Phys. B298 (1988) 741.
\bibitem{str}
 D. Garfinkle, G.~T. Horowitz and A. Strominger,
 Charged black holes in string theory,
 Phys. Rev. D43 (1991) 3140.
\bibitem{gal}
 E.~E. Donets and D.~V. Gal'tsov,
 Charged stringy black holes with non-abelian hair,
 Phys. Lett. B312 (1993) 391.
\bibitem{we}
 B. Kleihaus, J. Kunz and A. Sood, in preparation.
\bibitem{strau3}
 O. Brodbeck and N. Straumann,
 Instability proof for Einstein-Yang-Mills solitons
 and black holes with arbitrary gauge groups,
 ZU-TH-38-94, gr-qc/9411058.
\bibitem{volkov3}
 D.~V. Gal'tsov and M.~S. Volkov,
 Charged non-abelian SU(3) Einstein-Yang-Mills black holes,
 Phys. Lett. B274 (1992) 173.
\end{thebibliography}
\end{document}